\documentclass[aps,twocolumn,superscriptaddress,longbibliography,floatfix,notitlepage]{revtex4-1}
\usepackage{framed}
\usepackage{bbm}
\usepackage{physics}
\usepackage{amsmath}
\usepackage{epsfig}
\usepackage{subfigure,mathrsfs}
\usepackage{array}
\usepackage{amssymb}
\usepackage{braket}
\usepackage{float}
\usepackage{lmodern,amssymb}
\usepackage{physics}
\usepackage[dvipsnames]{xcolor}
\usepackage{lipsum}

\newcommand{\beq}[0]{\begin{equation}}
\newcommand{\eeq}[0]{\end{equation}}

\def\be{\begin{equation}}
\def\ee{\end{equation}}
\def\bea{\begin{eqnarray}}
\def\eea{\end{eqnarray}}
\newcommand{\ba}{\begin{eqnarray}}
\newcommand{\ea}{\end{eqnarray}}
\usepackage{hyperref}

\usepackage[bb=boondox]{mathalfa}

\usepackage{mathtools}

\begin{document}
\title  {Fluctuation theorem for a generalized observable depends on arbitrary two quantum variables}

\title  {Fluctuations and Irreversibility: Historical and Modern Perspectives}

\author{Sounak Bandyopadhyay}
\affiliation{Indian Institute of Technology Kanpur, 
	Kanpur, Uttar Pradesh 208016, India}

\author{Arnab Ghosh}
\thanks{arnab@iitk.ac.in}
\affiliation{Indian Institute of Technology Kanpur, 
	Kanpur, Uttar Pradesh 208016, India}

\begin{abstract}
		\indent This article traces the development of fluctuation theory and its deep connection to irreversibility, from equilibrium to near-equilibrium, and finally to far-from-equilibrium systems. Classical fluctuation theorems, which capture the statistical behaviour of thermodynamic systems far from equilibrium, are now well established. Their quantum counterparts, however, remain an active area of research. In this review, we highlight recent advances by linking quantum fluctuation theorems with linear response theory, offering new insights into the nature of quantum fluctuations and irreversibility in the near-equilibrium regime. Particular emphasis is placed on dissipated work in quantum systems as a pathway to observing non-classical effects in quantum thermodynamics. Understanding quantum fluctuations is not only essential for clarifying the foundations of irreversibility but also crucial for the development of novel quantum technologies, including quantum computers, sensors, and metrological devices.
		
		\end{abstract}
		%%insert keywords separated by semicolon using \keywords{words}
		
\maketitle

	%%close the twocolumn escape here
	%%include \shortTitle for the running head title (odd page)
	%%include \shortAuthor for the author running head (even page)
	%%include \doinum{number}for the DOI number in the header
	%%include \setcounter{page}{pagenum} for the exact starting page of the article. To be modified by Editorial office
	%%include \volnum{number} for the volume number in the header
	%%include \issuenum{number} for the volume number in the header
	%%include \year{yyyy} for  year of publication in the header
	%%include \pgrange{num--num} page range of article in the header
	
	\newcommand{\dbar}{d\hspace*{-0.08em}\bar{}\hspace*{0.1em}}

	\section{Introduction}

\begin{framed}
\noindent
``Thermodynamics is the only physical theory of universal content
which I am convinced will never be overthrown.''

\vspace{2pt}
\hfill --- Albert Einstein~\cite{einstein1926}
\end{framed}

%{\fontsize{27}{48}\calligra L} ightning strikes in a dense woodland, and forest fire breaks out. Predatory birds, animals were watchful of this opportunity, and others were afraid, but thus early humans came into contact with the thermal radiation caused by flames around 1-1.5 million years ago (Old Stone Age ~Palaeolithic). The relation of fire and human civilisation was regarded as the greatest invention by Darwin, excepting only language, even highlighted in various ancient cultures and mythologies. The fire eventually reshaped the human diet, resulting in an increase in brain size, which makes us wonder with curiosity about the manifestation of the universe.
	
	%{\fontsize{27}{48}\calligra T} 
	
	%Early humans came into contact with fire probably due to the forest fire breakouts ignited by lightning strikes around 1-1.5 million years ago (Old Stone Age $\sim$ Palaeolithic)~\cite{gowlett2016fire}. The relation of fire and human civilisation was regarded as the greatest invention by Darwin. The fire eventually reshaped the human diet, resulting in an increase in brain size, which makes us wonder with curiosity about the manifestation of the universe, even resonating with culture and mythology. From the middle (mesolithic) to the end (neolithic) of the Stone Age, sometimes between 5000 BCE, man found out that the natural circularity of tree trunks rolls more easily under gravity~\cite{rao2011history}. These rollers eventually developed into the wheels used in pottery, moving heavy objects, etc., which was a pivotal moment in human history~\cite{lee2024wheel}. 
	
	The control of \textit{fire} and the mastery of the \textit{wheel} have been regarded as two of humanity's greatest inventions. Even today, it is difficult to imagine building any complex structure without the wheel or the underlying principle of circular motion on an axis. However, it was not until 1763 that, while repairing Newcomen engines, James Watt combined the concepts of fire and the wheel to create the steam engine~\cite{rao2011history}. This innovation harnessed the energy of coal flames to produce rotary motion, powering locomotives, factory machinery and setting the stage for the Industrial Revolution. To pursuit the desire to optimise these widely used \textit{heat engines}, a remarkable group of people such as Sadi Carnot, James Joule, Rudolf Clausius, William Thompson, James Clerk Maxwell and Willard Gibbs laid the foundations of \textit{thermodynamics}~\cite{muller2007history}. Though initially formulated for equilibrium and reversible processes, thermodynamics eventually came to incorporate the crucial roles of \textit{fluctuations} and \textit{irreversibility} --- a topic of the present article. From the 1900s onwards, we further unveiled the atomic and molecular world, giving rise to \textit{quantum mechanics}. More recently, our curiosity has turned to the question: \textit{ `Are quantum thermodynamic machines better than their classical counterparts?'~\cite{ghosh2019are}} Just as the combination of fire and the wheel marked the beginning of the technological revolution, the harnessing of quantum resources may herald a new era. This review aims to briefly trace that remarkable journey.

	\indent The organisation of the paper is as follows: Section 2 presents a review of fluctuations in classical equilibrium and near-equilibrium thermodynamics. Section 3 introduces stochastic thermodynamics, where we discuss the various classical fluctuation theorems developed in the 1990s. Extensions to quantum fluctuation relations within the linear response regime are presented in Section 4. Finally, Section 5 concludes the paper.
	\section{Fluctuations in Equilibrium and Near-equilibrium Thermodynamics}

\begin{framed}
\noindent
``If your theory is found to be against the second law of thermodynamics
I can give you no hope.''

\vspace{2pt}
\hfill --- Arthur Eddington~\cite{eddington1928book}
\end{framed}

	From the middle of the twentieth century, our understanding of thermodynamics has changed dramatically. While classical thermodynamics relies on equilibrium properties, increasing importance is now given towards the study of fluctuations, instabilities, and evolutionary processes at all levels, from chemistry, biology to cosmology~\cite{kondepudi2015book}. Almost all processes in nature are irreversible, in which time-reversal symmetry is broken. The distinction between reversible and irreversible processes was introduced into thermodynamics through the concept of ``entropy'' --- the arrow of time, as Arthur Eddington called it~\cite{eddington1928book}. Unlike Newtonian mechanics, which originated in the study of celestial bodies, equilibrium thermodynamics --- as emphasised in the introduction --- arose during the Industrial Revolution as a phenomenological theory of macroscopic objects with a much more practical purpose: replacing manual labour with efficient machines by extracting \textit{useful work} from the \textit{motive power of fire}~\cite{carnot1824book}. To understand the thermodynamic behaviour of a system, it is therefore imperative to examine how it exchanges energy with its surroundings in the form of \textit{heat} and \textit{work}. Thus, the foundation of classical thermodynamics rests on two main pillars: \textit{energy} and \textit{entropy}~\cite{callen1985book}. Accordingly, thermodynamic systems are classified into three categories: isolated, closed, and open. In doing so, thermodynamics introduces the notion of \textit{temperature} --- a concept which was absent from classical mechanics. If the temperature of a system is not uniform everywhere, heat will flow until the entire system reaches a single temperature. Such a macroscopic state, characterized by the absence of change on the average, is called an \textit{equilibrium} state of the system. In particular, traditional thermodynamics focuses on average properties of macroscopic systems and disregards fluctuations. Since thermodynamic systems are typically large, thermal fluctuations are negligible; however, as the system size decreases, both thermal as well as quantum fluctuations become inevitable and must be taken into account~\cite{binder2019thermodynamics}. So, the description of non-equilibrium process, requires not only the averages but also the deviations of various thermodynamic quantities from its equilibrium values.  
	
	%\indent Historically, an irreversible transformation is usually defined by Clausius' inequality. 
	%\begin{equation}
	%    dS\ge \frac{dQ}{T}
	%\end{equation}
	In its original form, an irreversible transformation was characterized by Clausius~\cite{clausius1891book} in terms of 
	\begin{equation}
		N = S - S_0 - \int \frac{dQ}{T},
	\end{equation}
	where $S$ is the entropy of the final state and $S_0$ is the entropy of the initial state. According to him, ``The magnitude $N$ thus determines the \textit{uncompensated transformation} (\textit{uncompensirte Verwandlung})''~\cite{clausius1879book}. It represents the entropy produced by irreversible processes, which, by virtue of the second law of thermodynamics, can only be positive~\cite{prigogine1961book}. This production of entropy within the system establishes an arrow of time, distinguishing the future from the past. This stands in sharp contrast to the laws of mechanics --- both classical and quantum --- which are time symmetric and possess no intrinsic time-irreversibility in their fundamental equations. Thus, processes ruled out by the second law of thermodynamics, such as the spontaneous flow of gas molecules from a region of lower concentration to one of higher concentration, don't contradict with the laws of mechanics~\cite{kondepudi2015book}. This leads to a fundamental question: How can irreversible macroscopic processes, such as the flow of heat from higher to lower temperatures, emerge from the reversible microscopic dynamics of atoms and molecules? The answer lies in the foundations of statistical mechanics, which connect the probabilistic description of microscopic states with the macroscopic behaviour of thermodynamic systems~\cite{evans2016fundamentals}.

	%All irreversible processes, such as the flow of heat, arise from the motion of atoms and molecules governed by the laws of mechanics, which themselves make no distinction between evolution into the future and evolution into the past. Yet processes deemed impossible by the second law of thermodynamics do not violate the underlying mechanical laws.   

	%While, energy of a macroscopic system can be divided into kinetic and potential energies of a smaller systems, what about entropy? 

	To address these pertinent questions, Ludwig Boltzmann was first to introduce an extraordinary relation --- engraved on his memorial in the Central Cemetery in Vienna --- that entropy is a logarithmic measure of the number of microstates:~\cite{lebowitz1993boltzmann}  
	\begin{equation}
		S = k_B \ln W.
	\end{equation}
	Often, $W$ is referred to as the thermodynamic probability corresponding to the macrostate having entropy,~$S$. Now, the random motion of atoms and molecules causes all thermodynamic quantities to fluctuate around their equilibrium values. If the system slightly deviates from its equilibrium value $S_0$, we may express entropy as a function of a thermodynamic variable $A_i$ (such as internal energy, volume, or number of particles), while keeping all other parameters $A_{j \neq i}$ fixed:  
	\begin{equation}
		S = f(A_i).
	\end{equation}
	Expanding $S$ in a Taylor series around the equilibrium value $A_i^0$ and retaining terms up to second order, we obtain~\cite{kondepudi2015book}
	\begin{equation}
		S = S_0 + \bigg(\frac{\partial S}{\partial A_i}\bigg)_{A_i=A_i^0}(A_i - A_i^0) + \frac{1}{2}\bigg(\frac{\partial^2 S}{\partial A_i^2}\bigg)_{A_i=A_i^0}(A_i - A_i^0)^2 + \dots
	\end{equation}
	Here, $A_i^0$ is the equilibrium value of the thermodynamic variable $A_i$. Since entropy reaches its maximum value at equilibrium, we have  
	\begin{equation}
		\bigg(\frac{\partial S}{\partial A_i}\bigg)_{A_i=A_i^0} = 0,  
		\qquad \text{and} \qquad
		\bigg(\frac{\partial^2 S}{\partial A_i^2}\bigg)_{A_i=A_i^0} < 0.  
	\end{equation}
	Hence, the deviation of entropy from its equilibrium value due to a fluctuation in $A_i$ can be written as  
	\begin{equation}
		\Delta S = S - S_0 = \frac{1}{2}\bigg(\frac{\partial^2 S}{\partial A_i^2}\bigg)_{A_i=A_i^0}(A_i - A_i^0)^2,
	\end{equation}
	which is strictly negative. Thus, any fluctuation can only reduce the entropy of a system in equilibrium.

	Later, Albert Einstein prescribed a formula by rewriting Boltzmann's idea in reverse:~\cite{sommerfeld1964book}  
	\begin{equation}
		W = e^{S/k_B},
	\end{equation}
	where $S$ is considered to be known empirically, and $W$ is the unknown to be determined using the above relation. From this, we obtain a probability distribution function as a ratio of $W/W_0$:  
	\begin{equation}
		P(A_i) = \frac{W}{W_0} = e^{\frac{\Delta S}{k_B}} = \frac{1}{\sqrt{2\pi\sigma^2}} \exp\bigg[-\frac{(A_i - A_i^0)^2}{2\sigma^2}\bigg],
	\end{equation}
	where $W$ and $W_0$ are the thermodynamic probabilities corresponding to the entropies $S$ and $S_0$, respectively. This is a normalized Gaussian equilibrium probability distribution function, consistent with the central limit theorem, where $\sigma^2$ is the variance given by  
	\begin{equation}
		\sigma^2 = \langle(A_i - A_i^0)^2\rangle = -\frac{k_B}{\bigg(\frac{\partial^2 S}{\partial A_i^2}\bigg)_{A_i = A_i^0}},
	\end{equation}
	and is positive definite~\cite{kondepudi2015book}. As an example, if we calculate the fluctuation in volume for an ideal gas~\cite{dsray2015note}, the relative fluctuation tends to zero for large $N$:
	\begin{equation}
		\frac{\sigma^2}{\langle V \rangle^2} = \frac{\langle (V - \langle V \rangle)^2 \rangle}{\langle V \rangle^2} = \frac{1}{N} \longrightarrow 0 \quad (\text{for } N \to \infty),
	\end{equation}
	i.e., equilibrium fluctuations become negligible for a macroscopically large thermodynamic system.

	\indent Now, for multiple variables, we introduce the \textit{Onsager coordinates} $\alpha_i=A_i - A_i^0$ $(i=1,2,...)$, which represent the fluctuations of different thermodynamic variables around their equilibrium values. Proceeding as before, we expand the entropy, $S=f(\{A_i\})$, up to linear-order terms in $\alpha_i$s':~\cite{kondepudi2015book,dsray2015note}
	\begin{equation}
		S = S_0 + \sum_i \left(\frac{\partial S}{\partial \alpha_i}\right)_{\alpha_i=0}\alpha_i 
		+ \frac{1}{2}\sum_{ij}\left(\frac{\partial^2 S}{\partial \alpha_i \partial \alpha_j}\right)_{\alpha_i,\alpha_j=0}\alpha_i\alpha_j + \dots
	\end{equation}
	The change in entropy due to fluctuations is then
	\begin{equation}
		\Delta S = S - S_0 = \frac{1}{2}\sum_{ij}\left(\frac{\partial^2 S}{\partial \alpha_i \partial \alpha_j}\right)_{\alpha_i,\alpha_j=0}\alpha_i \alpha_j
	\end{equation}
	Since the system is in equilibrium, this change is negative, as before. So, the equilibrium state is stable against any perturbation that decreases entropy. Conversely, if fluctuation grows, the system is not in equilibrium. This principle is known as \textit{Gibbs stability theory}~\cite{kondepudi2015book}.

	\indent In response to a fluctuation that decreases entropy from its maximum equilibrium value, there will be irreversible processes producing entropy ($\Delta_i S$) that spontaneously drive the system back to equilibrium~\cite{kondepudi2015book}.  At equilibrium, these processes vanish and 19th-century thermodynamics was largely focused on such idealised reversible transformations. Recognising the connection between entropy and irreversibility, Pierre Duhem began developing a formalism that was later completed by Prigogine~\cite{prigogine1961book} and Onsager~\cite{onsager1931reciprocal-I,onsager1931reciprocal-II}. In the modern approach, entropy change is calculated in terms of variables that characterise non-equilibrium processes. Even if a system is not in \textit{global equilibrium}, thermodynamic quantities such as temperature, concentration, pressure, and internal energy remain well-defined locally, i.e., intensive parameters (temperature, pressure, etc.) are well defined within small volume elements, while extensive variables (entropy, internal energy, etc.) are expressed in terms of their corresponding densities. This is the assumption of \textit{local equilibrium}, which holds pretty well for most physical and chemical  systems.~\cite{prigogine1961book,kondepudi2015book}

	\indent Modern formalism due to Prigogine, begins with the infinitesimal change in entropy, written as~\cite{prigogine1961book,kondepudi2015book}  
	\begin{equation}
		dS = d_eS + d_iS ,
	\end{equation}
	where $d_eS$ is the entropy change due to the exchange of matter and energy with the environment, and $d_iS$ is the entropy change due to the \textit{uncompensated transformation}, i.e., the entropy produced by irreversible transformations. This in turn can be described by the sum of individual thermodynamic forces (such as gradients of temperature, concentration, etc.) and their corresponding thermodynamic fluxes (such as heat flow, diffusion, etc.). The rate of entropy production can thus be expressed as~\cite{callen1985book, kondepudi2015book}
	\begin{equation}\label{sigma-dot}
		\dot{\Sigma} = \frac{d_iS}{dt} = \sum_k F_k J_k ,
	\end{equation}
	where $F_k$ and $J_k$ are the respective thermodynamic forces and fluxes. The entropy production rate and the entropy production itself are always positive ($\dot{\Sigma} > 0,\; d_iS > 0$). This statement is stronger and more general than the classical formulation that the entropy of an isolated system can only increase. This insight by Prigogine earned him the Nobel Prize in Chemistry in 1977~\cite{prigogine1977time}.

	\indent To obtain Eq.~\eqref{sigma-dot} for the $\dot{\Sigma}$ associated with fluctuations in system variables, the thermodynamic force can be written as, $F_k = \frac{\partial \Delta_i S}{\partial \alpha_k}$, conjugate to the thermodynamic flux $J_k = \frac{\partial \alpha_k}{\partial t}$. If the deviations of the forces from their equilibrium values are small, the fluxes can be expressed as \textit{linear} functions of the forces, $J_k = \sum_j L_{kj} F_j$,
	where $L_{kj}$ are constants known as \textit{kinetic coefficients}. Thus, the rate of entropy production can be calculated as~\cite{kondepudi2015book}
	\begin{eqnarray}
		\dot{\Sigma} = \frac{\partial \Delta_i S}{\partial t} 
		&=& \sum_k \frac{\partial \Delta_i S}{\partial \alpha_k} \frac{\partial \alpha_k}{\partial t} \nonumber \\
		&=& \sum_k F_k J_k = \sum_{jk} F_j L_{jk} F_k > 0.
	\end{eqnarray}
	Here, $\mathbf{L}$ is a positive definite matrix whose elements satisfy the symmetry condition $L_{kj} = L_{jk}$. These reciprocal relations are linked to thermodynamic cross-effects and were first observed by Lord Kelvin and others in the study of thermoelectric phenomena (such as the Seebeck (1821) and Peltier (1834) effects in metal wires)~\cite{kondepudi2015book,gupt2024graph}. At that time, they were regarded merely as conjectures. A rigorous mathematical derivation was later provided by Lars Onsager in 1931, after which they became known as the \textit{Onsager reciprocal relations}~\cite{onsager1931reciprocal-I,onsager1931reciprocal-II}. The foundation of Onsager’s theory rests on the properties of steady-state fluctuations:
	\begin{equation}
		\langle \alpha_j(t)\alpha_k(t+\tau)\rangle 
		= \langle \alpha_j(t)\alpha_k(t-\tau)\rangle
		= \langle \alpha_j(t+\tau)\alpha_k(t)\rangle .
	\end{equation}
	The first equality reflects time-reversal symmetry, while the second expresses time-translation invariance. Taken together, these properties --- often referred to as \textit{microscopic reversibility} or the \textit{principle of detailed balance} --- lead directly to the reciprocal relations $L_{kj} = L_{jk}$. The importance of this result is so profound that it is regarded as ``fourth law of thermodynamics,''~\cite{deffner2019quantumthermo} a contribution for which Onsager was awarded the Nobel Prize in Chemistry in 1968~\cite{onsager1968the}.

	We conclude this section by highlighting the deep connection between entropy and fluctuations in equilibrium and near-equilibrium regimes. This connection explains why equilibrium thermodynamic properties are primarily governed by average values and how microscopic fluctuation symmetries give rise to thermodynamic cross-effects at the macroscopic scale. To fully comprehend how irreversibility emerges at macroscopic length scales from time-reversible microscopic dynamics, one must turn to the field of \textit{stochastic thermodynamics}~\cite{evans2016fundamentals}, which came to the forefront in the 1990s.

	%\textcolor{red}{Eq. 14 is no longer in equilibrium; this is already within the linear response regime and near-equilibrium phenomenon. Think about it and change it accordingly!!}
	
\section{Emergence of Stochastic Thermodynamics and Fluctuation Theorems}

			\begin{framed}
\noindent
``If we shift our focus away from equilibrium states,
we find a rich universe of nonequilibrium behavior.''

\vspace{2pt}
\hfill --- Christopher Jarzynski~\cite{jarzynski2015diverse}
\end{framed}

	\indent Over the past three decades, stochastic thermodynamics has become a leading framework for understanding diverse non-equilibrium phenomena, ranging from molecular motors and cellular reactions to colloids and nanoscale systems~\cite{udo2012stochastic,peliti2021stochastic, limmer2024statistical, seifert2025stochastic}. Often, thermodynamics on this scale refers as \textit{nanoscale thermodynamics}~\cite{broedersz2022twenty}, and researchers are now attempting to extend this concept to the quantum realm, which will be discussed in the following section~\cite{benenti2017fundamental}. Broadly speaking, stochastic thermodynamics describes mesoscopic nonequilibrium systems in contact with equilibrium heat bath. Here, the system interacts randomly with the reservoir, so that at time $t$ it may be found in a state $x$ with probability $P(x;t)$. In such small systems, fluctuations play a pivotal role in controlling the dynamics of the system, and thermodynamic quantities such as heat, work, and entropy themselves fluctuate. Unlike in the classical case, these quantities don't take on single fixed values but are instead described by probability distributions, with mean values serving as the relevant averages. Remarkably, these distributions for thermodynamic observables fulfil some symmetry relations known as \textit{fluctuation theorems} (FTs)~\cite{seifert2025stochastic}. Here, we will discuss a few such relations.
	
	%While the laws of thermodynamics are universal, but depending on different scales  different approximations are applied and distinct insights are extracted about the system.

	\subsubsection{Evans-Searles fluctuation theorem (1994):}

	Evans and Searles proposed the first fluctuation relation related to general dissipation in simulations of sheared fluids~\cite{evans1994equilibrium}. Here, we present a general version of the theorem within Hamiltonian dynamics. Consider an isolated Hamiltonian system of interacting particles. The state of the system is represented by a phase-space vector containing the coordinates and canonical momenta of all particles, in an Avogadro-dimensional space, denoted as $\Gamma \equiv \{\textbf{q}_i,\textbf{p}_i\}$. The Hamilton's equations for such an autonomous system are given by
	\begin{eqnarray}\label{Hamilton's-eqn}
		\dot{\textbf{q}}_i &=& \frac{\partial H(\{\textbf{q}_i,\textbf{p}_i\})}{\partial \textbf{p}_i}, \qquad 
		\dot{\textbf{p}}_i = -\frac{\partial H(\{\textbf{q}_i,\textbf{p}_i\})}{\partial \textbf{q}_i}.    
	\end{eqnarray}
	%$\Gamma$ may include additional dynamical variables as extended phase space vector, such as the volume of the isobaric systems, or the thermostat multiplier associated with possible Nosé–Hoover thermostat.
	Due to the large number of degrees of freedom, it is not feasible to specify the initial condition of the system by explicitly providing all $\{\mathbf{q}_i(0),\mathbf{p}_i(0)\}$. Instead, we introduce the probability distribution function (PDF), $f(\Gamma,t)$, with the initial condition
	\begin{equation}
		f(\Gamma,0)=\frac{e^{-F(\Gamma)}}{\int_D d\Gamma \, e^{-F(\Gamma)}}.
	\end{equation}
	Here, $F(\Gamma)$ is an arbitrary single-valued real function defined over a specified phase-space domain $D$. Defining the PDF as $f(\Gamma,t)$, the probability of finding the system in an infinitesimal volume of phase-space $d\Gamma_t$ at time $t$ is given by $P(\Gamma_t,t)=f(\Gamma_t,t)\, d\Gamma_t$. Using the total time derivative of $f(\Gamma_t,t)$ and the continuity equation for probability flow in phase space, we obtain  
	\begin{equation}\label{eq:f(gamma,t)}
		\frac{df}{dt}=\frac{\partial f}{\partial t}+ \frac{\partial f}{\partial \Gamma}\dot{\Gamma}, 
		\qquad \text{and} \qquad
		\frac{\partial (f\dot{\Gamma})}{\partial \Gamma}+\frac{\partial f}{\partial t}=0.
	\end{equation}
	respectively. From Eq.~\eqref{eq:f(gamma,t)}, the time evolution of $f(\Gamma_t,t)$ can be calculated as follows:
	\begin{equation}\label{eq:df-dt}
		\frac{df}{dt}=-f\Lambda, \qquad \text{where} \quad \Lambda=\frac{\partial\dot{\Gamma}}{\partial\Gamma}.
	\end{equation}
	Integrating Eq.~\eqref{eq:df-dt} from $t=0$ to $t=\tau$, yields  
	\begin{equation}\label{soln-f-gamma-t}
		f(\Gamma_\tau,\tau)= f(\Gamma_0,0)\, e^{-\int_0^\tau \Lambda \, dt}.
	\end{equation}
	Since the probability $P(\Gamma_t,t)$ of a specific phase-space volume is conserved during evolution, we have 
	\begin{equation}\label{con-phase-sp-vol}
		f(\Gamma_{\tau},\tau)\, d\Gamma_{\tau} = f(\Gamma_{0},0)\, d\Gamma_{0}.
	\end{equation}
	Combining Eqs.~\eqref{soln-f-gamma-t}-\eqref{con-phase-sp-vol}, we find that the compression factor for the infinitesimal phase-space volume is given by 
	\begin{equation}\label{eq:compression-factor}
		\frac{d\Gamma_\tau}{d\Gamma_0}=e^{\int_0^\tau \Lambda \, dt},
	\end{equation}
	which plays a central role in formulating FTs.

	%

	%As we already mentioned, the PDF as $f(\Gamma,t)$, the probability of finding the system in an infinitesimal phase space volume $d\Gamma$ is $P(d\Gamma_t,t)=f(\Gamma_t,t)d \Gamma_t$.

	%so we make use of some known but arbitrary probability distribution function $f(\Gamma,t)$, which initially is

	\indent To understand the origin of the time arrow, we first note that the microscopic equations of motion, Eq.~\eqref{Hamilton's-eqn}, is invariant under time reversal. Since the dynamics here are deterministic, any trajectory is uniquely determined by its initial phase-space point. Let $S^{\tau}$ denote the time evolution operator that acts over the time interval $0$ to $\tau$. Thus, for a trajectory starting at $\Gamma_0$, we have $\Gamma_{\tau} = S^{\tau}\Gamma_0$. Now, let $M^{\tau}$ be the time-reversal mapping operator, which reverses the momenta as $M^{\tau}\Gamma=M^{\tau}\{\textbf{q}_i,\textbf{p}_i\}=\{\textbf{q}_i,-\textbf{p}_i\}$. So, for the ``\textit{anti-trajectory}'' conjugate to $\Gamma_0$, we can depict the initial phase-space point by $\Gamma^{*}_{0} = M^{\tau}\Gamma_{\tau} = M^{\tau}S^{\tau}\Gamma_0$. Since time reversal leaves the phase-space volume unchanged, we have $d\Gamma_0^* = d\Gamma_\tau$, which, upon using Eq.~\eqref{eq:compression-factor}, becomes
	\begin{equation}
		d\Gamma_0^* = d\Gamma_0 \, e^{\int_0^\tau \Lambda dt}.
		\label{eq:volume2}
	\end{equation}

	\indent Next, instead of a single trajectory, we consider a bundle of trajectories occupying a phase-space volume $d\Gamma_t$ with probability $P(d\Gamma_t,t)$. Microscopic reversibility requires equal probabilities for both forward and reverse bundles of trajectories, i.e., $P(d\Gamma_0,0)=P(d\Gamma_0^*,0)$, which leads to   
	\begin{equation}
		f(d\Gamma_0,0)\,d\Gamma_0 = f(d\Gamma_0^*,0)\,d\Gamma_0^* .  
		\label{eq:microrev}
	\end{equation}
	Since it is based on the \textit{principle of microscopic reversibility}, Eq.~\eqref{eq:microrev} could be regarded as the criterion for equilibrium. Combining it with Eq.~\eqref{eq:volume2} yields   
	\begin{equation}
		\frac{f(d\Gamma_0,0)}{f(d\Gamma_0^*,0)} = e^{\int_0^\tau \Lambda dt}.  
		\label{eq:ratio}
	\end{equation}
	If the system is not in equilibrium, then $P(d\Gamma_0,0) \neq P(d\Gamma_0^*,0)$. Hence, we can, in general, define a dissipation function along a trajectory originating from the phase-space point $\Gamma_0$ as
	\begin{equation}\label{dissipation-fn}
		\Omega_\tau(\Gamma_0)=\ln\bigg(\frac{P(d\Gamma_0,0)}{P(d\Gamma_0^*,0)}\bigg)=\ln\bigg(\frac{f(d\Gamma_0,0)}{f(d\Gamma_0^*,0)}\bigg)-\int^\tau_0\Lambda dt,
	\end{equation}
	which quantifies the deviation from equilibrium. For $\Omega_\tau(\Gamma_0)$ to be well defined over the phase space domain $D$, the system must be \textit{ergodically consistent} over $D$~\cite{evans2016fundamentals}. It is evident that $\Omega_\tau(\Gamma_0)=0$ at equilibrium. Since $\Omega_\tau(\Gamma_0)$ characterizes the reversibility of a set of trajectories, we can evaluate the relative probability of the dissipation function taking opposite values:
	\begin{equation}\label{Evans-Searles-FT}
		\frac{\mathcal{P}(\Omega_\tau=A)}{\mathcal{P}(\Omega_\tau=-A)}=\frac{\int d\Gamma_0\delta(\Omega_\tau(\Gamma_0)-A)f(\Gamma_0,0)}{\int d\Gamma_0^*\delta(\Omega_\tau(\Gamma_0^*)+A)f(\Gamma_0^*,0)} = e^A , 
	\end{equation}
	where we used the key property of the odd parity of the dissipation function, i.e., $\Omega_\tau(\Gamma_0)=-\Omega_\tau(\Gamma_0^{*})$, which follows directly from its definition in Eq.~\eqref{dissipation-fn}. Equation~\eqref{Evans-Searles-FT} is known as the \textit{Evans–Searles fluctuation theorem} (ESFT)~\cite{evans1994equilibrium}. This implies that dissipation functions taking positive values for \textit{forward} irreversible trajectories are more probable than for the corresponding \textit{reverse} antitrajectories. Thus, the dissipation function can be viewed as a measure of the temporal asymmetry inherent in bundles of trajectories originating from an initial distribution of states. In defining the dissipation function in Eq.~\eqref{dissipation-fn}, which characterizes deviations from equilibrium, we did not restrict ourselves to the near-equilibrium situation. As a consequence, the ESFT derived here applies to nonequilibrium fluctuations arbitrarily far from equilibrium and to systems of any size, without requiring the classical thermodynamic limit. Most importantly, the ESFT demonstrates that even deterministic, time-reversible equations of motion can give rise to irreversibility~\cite{evans2016fundamentals}. The ESFT has also been verified experimentally, for instance by Wang \textit{et al.} (2002)~\cite{wang2002experimental} and Reid \textit{et al.} (2004)~\cite{reid2004reversibility}.

	%In defining the dissipation function in Eq.~\eqref{dissipation-fn} characterizingdeviation from equilibrium, we have not restricted ourselves to near equilibrium situation. As a consequence, ESFT result derived here on non-equilibrium fluctuation valid arbitrary far from equilibrium and apply to a system of arbitrary size (no need to take classical thermodynamic limit). It implies that the dissipation function taking positive value for the \textit{forward} irreversible trajectories is more favourable than the \textit{reverse}  antitrajectories. Thus, one way to think of the dissipation function is as measure of the temporal asymmetry inherent in bundle of trajectories originating from an initial distribution of states. Most, importantly, ESFT shows even in a deterministic, time-revesible equations of motion can produce irreversibility. 

	One can further evaluate the phase-space average of the dissipation function as $\langle \exp(-\Omega_{\tau})\rangle=1$.  
	Since $\exp(-\Omega_{\tau})$ is a convex function, Jensen’s inequality gives~\cite{seifert2025stochastic}  
	\begin{equation}
		\exp(-\langle\Omega_{\tau} \rangle) \leq \langle \exp(-\Omega_{\tau})\rangle=1 ,
	\end{equation}
	which implies \(\langle \Omega_{\tau} \rangle \geq 0\). This shows that although the dissipation function defined in Eq.~\eqref{dissipation-fn} can take positive, negative, or zero values, its average is always nonnegative. This is the statement of the second law of thermodynamics, highlighting the link between microscopic reversible dynamics and macroscopic irreversible behavior, and providing a resolution of the Loschmidt paradox~\cite{evans2016fundamentals}. In the following, we extend the FTs to more practical thermodynamic quantities.

	\subsubsection{Jarzynski equality (1997):}

	Jarzynski developed a more practical relation that allows us to relate the work $W$ and the free energy change $\Delta F$ for a non-quasistatic process between two sates of equilibrium. It is an equality and hence contains more information than classical thermodynamics, which only connects them through an inequality $W \geq \Delta F$~\cite{jarzynski1997nonequilibrium,jarzynski2010equalities}.

	\indent Consider a driven isolated system with a $\lambda$-dependent Hamiltonian $H(\Gamma,\lambda)$ defined on the domain $D_\lambda$ of the phase space. At equilibrium, for each value of $\lambda$ and inverse temperature $\beta$, there corresponds a free energy 
	\begin{equation}
		F(\beta,\lambda)=-(1/\beta)\ln{\sum_{\Gamma\in D_\lambda}e^{-\beta H(\Gamma,\lambda)}}.
	\end{equation}
	The system is driven through a \textit{regular protocol} $\lambda(t)$, with $0\le t\le \tau$, from an initial value $\lambda(0)$ to a final value $\lambda(\tau)$. For an initial phase point $\Gamma_0$, the work applied up to time $\tau$ is defined as
	\begin{equation}
		\begin{split}
			W(\Gamma_0)&=H(\Gamma_{\tau}(\Gamma_0),\lambda(\tau))-H(\Gamma_0,\lambda(0))\\
			&=\int_0^\tau dt \partial_\lambda H(\Gamma_t,\lambda)\bigg|_{\lambda=\lambda(t)}\partial_t\lambda(t).
		\end{split}
	\end{equation}
	If the system is initially thermalised at inverse temperature $\beta$, the normalised initial canonical distribution $p_i(\Gamma_0)$ is given by
	\begin{equation}\label{eq:pixii}
		p_i(\Gamma_0)=e^{-\beta [H(\Gamma_0,\lambda(0))-F(\beta,\lambda(0))]}
	\end{equation}
	Now, for any auxiliary final distribution $p_f(\Gamma_1)$, normalised on $D_1$, we can write
	\begin{equation} \label{eq:p1xi1} p_f(\Gamma_1)=p_f(\Gamma_{\tau}(\Gamma_0))=e^{-\beta [H(\Gamma_{\tau}(\Gamma_0)),\lambda(\tau))-F(\beta,\lambda(\tau)]}
	\end{equation}
	Since, for a regular protocol, the dynamical image of $D_0$ after time $\tau$ becomes $D_1$, the normalisation of $p_f(\Gamma_1)$ is guaranteed. So, we can write~\cite{seifert2025stochastic}
	\begin{equation}\label{eq:IFT}
		\begin{split}
			\sum_{\Gamma_1\in D_1}p_f(\Gamma_1)= 1=& \sum_{\Gamma_0\in D_0}p_i(\Gamma_0)\frac{p_f(\Gamma_{\tau}(\Gamma_0))}{p_i(\Gamma_0)}\\
			=& ~\bigg\langle\frac{p_f(\Gamma_{\tau}(\Gamma_0))}{p_i(\Gamma_0)}\bigg\rangle=\langle e^{-\Omega(\Gamma_0)}\rangle
		\end{split}
	\end{equation}
	This remarkable property, which applies for the distribution of work and several other thermodynamic quantities, is known as the \textit{integral fluctuation theorem} (IFT)~\cite{udo2012stochastic, seifert2025stochastic}. From Eqs.~\eqref{eq:pixii}-\eqref{eq:IFT}, we identify
	\begin{equation}
		\begin{split}\label{eq:R-xi-0}
			\Omega(\Gamma_0)=&\beta[H(\Gamma_{\tau}(\Gamma_0),\lambda(\tau))-H(\Gamma_0,\lambda(0))-\Delta F]\\
			=&\beta[W(\Gamma_0)-\Delta F]
		\end{split}
	\end{equation}
	with $\Delta F = F(\beta,\lambda(\tau))-F(\beta,\lambda(0))$. Substituting Eq.~\eqref{eq:R-xi-0} into Eq.~\eqref{eq:IFT}, the corresponding IFT reduces to the celebrated \textit{Jarzynski equality} (JE)~\cite{jarzynski1997nonequilibrium}:
	\begin{equation}\label{Jarzynski-Equality}
		\langle \exp(-\beta W(\Gamma_0))\rangle=\exp(-\beta\Delta F).
	\end{equation}
	
	\indent This equality relates an average over the work required in this process to the free-energy difference between the two canonical equilibrium states: one corresponding to the value $\lambda(\tau)$ of the control parameter at time $\tau$ and other to its initial value $\lambda(0)$, both taken w.r.t. the same inverse temperature $\beta$. Further, applying Jensen's inequality to Eq.~\eqref{Jarzynski-Equality}, gives $\mathcal{W}=\langle W(\Gamma_0)\rangle\ge \Delta F$~\cite{udo2012stochastic}. If we define the dissipative work as $W_{\text{diss}}= W-\Delta F$, this inequality can be rewritten as $\langle W_{\text{diss}}\rangle\ge\ 0$~\cite{seifert2025stochastic}. For smaller systems, unlike in the thermodynamic limit, $W_{\text{diss}}$ need not always be positive for individual realisations of the process; it may even be negative. Nevertheless, upon averaging, it becomes consistent with classical thermodynamics. If we identify $W_{\text{diss}}=T \Sigma$, where $\Sigma$ is the entropy production, then combining the IFT with Jensen's inequality leads to $\langle \Sigma(\Gamma)\rangle\ge0$ which is recognised as the \textit{second law of stochastic thermodynamics}~\cite{peliti2021stochastic}. The application of JE to mesoscopic systems in modern times has been remarkable. Notably, JE places no restriction on the speed of the transformation and thus remains valid even far from equilibrium. This makes it a powerful tool for determining free-energy differences in molecular dynamics simulations, driven harmonic oscillators, and even real biomolecular experiments~\cite{broedersz2022twenty}. Many researchers view JE as having brought \textit{nineteenth-century thermodynamics} into the \textit{twenty-first century}, given its profound implications~\cite{deffner2019quantumthermo}.

	%\textcolor{red}{Be careful, the definition of $S$ in $W_{diss}=\beta^{-1} S$ and in Sect. 2 is different. This happens when you take materials from different sources! }

	%\textcolor{red}{Is your dynamical image $D_\lambda$ same as the phase space volume $\Gamma$ used in the previous section? If yes, it is always better to use uniform notation throughout one text. One should not change notation between different sections}

	\subsubsection{Crooks Theorem (1999):}

	Although formulated after Jarzynski’s work, Crooks’ theorem provides a more general framework, with the JE arising as a direct consequence. It also illustrates the origin of the arrow of time by introducing the distribution of work: $\mathcal{P}_F(W)$ for the \textit{forward process}, governed by the protocol $\lambda(t)$, and $\mathcal{P}_R(-W)$ for the \textit{reverse process}, governed by the protocol $\lambda(\tau-t)$. Both processes start from thermal equilibrium at inverse temperature $\beta$. Accordingly, we obtain  
	\begin{equation}
		\begin{split}
			&\mathcal{P}_R(-W)=\\
            =&\sum_{\Gamma^R_0\in D_R}\delta(W_R(\Gamma^R_0)+W)e^{-\beta[H(\Gamma^R_0,\lambda(\tau))-F(\beta,\lambda(\tau))]}\\
			=&\sum_{\Gamma_0\in D_F}\delta(-W_F(\Gamma_0)+W)e^{-\beta[H(\Gamma_0,\lambda(0))+W_F(\Gamma_0)-F(\beta,\lambda(\tau))]}\\
			=& \mathcal{P}_F(W)e^{-\beta(W-\Delta F)}
		\end{split}
	\end{equation}
	%\textcolor{red}{I think still there are errors! Please check carefully! The brackets are not closed and many more!}\textcolor{blue}{ok actually it is from the Seifert book I will check again}
	Finally, we can write~\cite{crooks1998nonequilibrium,crooks1999entropy}
	\begin{equation}\label{eqf:CFT}
		\frac{\mathcal{P}_F(W)}{\mathcal{P}_R(-W)}=e^{\beta(W-\Delta F)}.
	\end{equation}
	This is the renowned \textit{Crooks fluctuation theorem} (CFT), which shows that the reverse process is exponentially suppressed, giving rise to the direction of time, much like the second law of thermodynamics. In the derivation, we use the antisymmetry of work-energy conservation and the fact that momenta appear quadratically in the Hamiltonian, i.e.,
	\begin{equation}
		\begin{split}
			H(\Gamma_0^R,\lambda(\tau))=&H(M^\tau \Gamma_{\tau},\lambda(\tau))=H(\Gamma_{\tau},\lambda(\tau))\\=&H(\Gamma_0,\lambda(0))+W(\Gamma_0).
		\end{split}
	\end{equation}
	The CFT relation in Eq.~\eqref{eqf:CFT} is stronger than JE [cf. Eq.~\eqref{Jarzynski-Equality}], since JE can be derived by integrating both sides of Eq.~\eqref{eqf:CFT} and using $\int dW \mathcal{P}_R(-W)=1$. Collin \textit{et al.} (2005)~\cite{collin2005verification} experimentally verified Crooks’ theorem in a single-molecule RNA hairpin folding–unfolding experiment using optical tweezers. In this experiment, $W_{\text{diss}}$ was sometimes negative, but on average it was positive. Thus, for small biological systems, this theorem provides a clear-cut framework for calculating free-energy changes from work distributions.

	\subsubsection{The different DFT and furthermore:}
	
	\indent If we consider symmetric driving protocol, $\lambda(t)=\lambda(\tau-t)$, then the work distributions for the forward and backward processes become identical and $\Delta F=0$. In this case, the \textit{Crooks theorem} reduces to
	\begin{equation}
		\frac{\mathcal{P}(W)}{\mathcal{P}(-W)}=e^{\beta W},
	\end{equation}
	which is an example of the \textit{detailed fluctuation theorem} (DFT)~\cite{seifert2025stochastic}. The DFT is a stronger result than the IFT. While the IFT provides a single global constraint on the work distribution (and more generally on entropy production), the DFT establishes a local balance relation between the probabilities of observing a trajectory and its corresponding antitrajectory. Consequently, in a steady-state system, it is sufficient to know $\mathcal{P}(W)$ for positive values of $W$ in order to reconstruct the entire distribution~\cite{peliti2021stochastic}. Both the JE and the CFT can be regarded as special cases of the IFT and DFT when the initial and final states are equilibrium states.

	There exists a vast literature on FTs, depending on the observables of interest, the underlying dynamics of the system, and the hypotheses or models considered. Evans and Searles~\cite{evans1994equilibrium} first proposed such a relation. Subsequently, Gallavotti and Cohen~\cite{gallavotti1995dynamical} derived a detailed fluctuation theorem for the entropy production rate in deterministic chaotic systems. Later, the JE~\cite{jarzynski1997nonequilibrium} and CFT~\cite{crooks1999entropy} were developed for systems driven between two equilibrium states, becoming landmark results that enabled equilibrium free-energy differences to be estimated from non-equilibrium measurements. This also led to the rediscovery of earlier works by Bochkov and Kuzovlev~\cite{bochkov1977general}, although their definition of work differs from the present one.

	\indent Moreover, introduction of trajectory probabilities for forward and backward processes naturally gave rise to a variety of FTs. For entropy production, if we separate its contributions into two parts --- one arising from time dependence and another from external driving (adiabatic and non-adiabatic entropy production) --- the IFT leads to the \textit{Hatano-Sasa theorem}~\cite{hatano2001steady}, which constitutes a cornerstone in the generalisation of FDR (\textit{fluctuation-dissipation relation}) to \textit{non-equilibrium steady state}. Subsequently, Rao and Esposito~\cite{rao2018detailed} developed a unified framework that systematically derives many of the FTs reported in the literature. In addition to work and entropy, considerable attention has been devoted to the fluctuation symmetry of heat~\cite{baiesi2006fluctuation,noh2012fluctuation}. Jarzynski and Wójcik~\cite{jarzynski2004classical} analysed the statistics of heat exchange between two systems coupled to distinct thermal baths, deriving a fluctuation relation that is now widely referred as \textit{exchange fluctuation theorem} (XFT/EFT)~\cite{timpanaro2019thermodynamic}. More recently, the study of joint statistics of thermodynamic quantities and their associated FTs has attracted considerable attention. Interestingly, while the individual quantities may or may not obey an FT on their own, their \textit{joint probability distribution} often satisfies a \textit{joint fluctuation theorem}, thereby revealing richer symmetry structures in far from equilibrium thermodynamics~\cite{miller2021joint,chen2023hierarchical}.

	So far, we have provided an account of the development of FTs in a classical settings. To understand their quantum counterparts, however, we must dig deeper. The key difference arises because any measurement of quantum states or observables disturbs the system and its trajectories, unlike in classical systems. In fact, continuous monitoring can even freeze the dynamics due to the \textit{quantum Zeno effect}~\cite{aulettaparisi2009quantum}. In the quantum regime, the very definitions of work, heat, and entropy must be handled with care. Thus, exploring quantum versions of FTs is a rich and promising area of current research. 
	
	%In the final section of this review, we will shed light on this from the perspective of linear response theory. 

	%\indent So far, we gave a vivid understanding of the development of FTs in a classical deterministic world, but to understand their quantum version, we have to dig deeper. The key difference comes because any measurement of any quantum states or observables disturbs the system or trajectories, unlike classical systems, since continuous monitoring of the system would freeze the dynamics according to \textit{quantum zeno effect}. In the quantum world, the definition of work, heat, and even entropy deserves to be handled with care. So, understanding these quantum versions of FTs is a promising field of research, which will be discussed in the later sections.
	
	\section{Linear Response Theory and Quantum Work Fluctuations}

				\begin{framed}
\noindent
``The idea of dissipation of energy depends on the extent
of our knowledge.''

\vspace{2pt}
\hfill --- James Clerk Maxwell~\cite{maxwell1890the}
\end{framed}

	%\begin{center}
	%\begin{minipage}{\textwidth}
	%\begin{tcolorbox}[colback=gray!10, colframe=black!50, width=0.48\textwidth, boxrule=0.5pt, arc=3mm]
	%\itshape
	%``LRT provides a general and unified method of treating the irreversible processes, valid for a wide class of systems irrespective of their microscopic details.''  
	
	%\hfill --- Ryogo Kubo
	%\end{tcolorbox}
	%\end{minipage}
	%\end{center}

	Up to this point, we have discussed the role of fluctuations in a thermodynamic setup, where general thermodynamic quantities vary due to interactions with heat bath. In 1957, Ryogo Kubo~\cite{kubo1957statistical} laid a concrete foundation within the Hamiltonian framework, showing how any arbitrary dynamical observable of a system coupled to a thermal bath deviates from equilibrium under an external drive, and how irreversible macroscopic behavior appears as a direct consequence of microscopic fluctuations~\cite{kubo1966thefluctuation} when observed through the lens of \textit{linear response} of the system. Since then, Linear Response Theory (LRT)~\cite{kubo1983book,zwanzig2001nonequilibrium,pottier2010nonequilibrium,balakrishnan2020elements} has held a central place as a powerful tool for studying near-equilibrium systems, with far-reaching applications in fluid dynamics, transport phenomena (both classical and quantum), condensed matter physics, many-body problems, chemical kinetics, and beyond~\cite{nitzan2024chemical}. More recently, it started elucidating its impact on \textit{quantum thermodynamics} (QTD)~\cite{guarnieri2024generalized,gupt2024graph}.
	
	%Up to this point, we have discussed the impact of fluctuations in a thermodynamic setup where the general thermodynamic quantities vary due to interaction with a heat bath noise. Ryogo Kubo in the year of 1957 provides a concrete foundation in the Hamiltonian framework that how any arbitrary dynamic observable of a system, connected to a thermal bath, deviates from its equilibrium due to an external drive and how that new average (away from equilibrium) can be expressed in terms of \textit{fluctuations} in equilibrium quantities. After the discovery, LRT has maintained its pivotal position as a tool to explore near equilibrium systems with paramount applications in the  field of fluid dynamics, transport phenomena (classical and quantum both), condensed matter physics, many body problems,  chemical kinetics and many more. Now, it is also elucidating its impact in \textit{quantum thermodynamics}.

	\subsubsection{Introduction to LRT:}
	Here, we establish the Kubo formulas in parallel for both classical~\cite{balakrishnan2020elements,balakrishnan2020mathematical,kubo1983book,pottier2010nonequilibrium,zwanzig2001nonequilibrium} and quantum mechanics~\cite{kubo1983book,pottier2010nonequilibrium}. These formulas describe correlation of fluctuations, such as response and relaxation functions between observables of different dynamical variables. We discuss the properties and applications of these response and relaxation functions in the context of QTD.

	\indent Consider a physical system (more precisely, a subsystem of interest) in contact with a heat bath at inverse temperature $\beta = 1/k_B T$. Initially, the system is governed by a time-independent Hamiltonian $H_0(\{\mathbf{q}_i,\mathbf{p}_i\})$, 
	which describes its natural dynamics. The normalized canonical equilibrium phase-space density (or, equilibrium density operator in case of quantum mechanics) is given by  
	\begin{equation}
		\rho_{\text{eq}} = \frac{e^{-\beta H_0}}{\mathrm{Tr}[e^{-\beta H_0}]} .
	\end{equation}
	The equilibrium average of an arbitrary observable $B(q,p)$ is then  
	\begin{equation}\label{eqm_B}
		\langle B \rangle_{\rho_{\text{eq}}} 
		= \frac{\mathrm{Tr}[\rho_{\text{eq}} B]}{\mathrm{Tr}[\rho_{\text{eq}}]} 
		= \frac{\int dq \int dp \, B(q,p)\, e^{-\beta H_0(q,p)}}{\int dq \int dp \, e^{-\beta H_0(q,p)}} .
	\end{equation}
	We now perturb the system, starting from an initial time $t_0$ (eventually we take $t_0\to-\infty$) by applying a time-dependent external force $\lambda(t)$. This force is spatially homogeneous and couples with an arbitrary observable $A$ of the system. The perturbed Hamiltonian is then $H(t)=H_0+H'(t)=H_0-\lambda(t)A$. For $t\ge t_0$, the density operator (or the phase-space distribution in the classical case), $\rho(t)$ evolves according to the Liouville–von Neumann equation~\cite{breuer2002book,nitzan2024chemical}
	\begin{equation}\label{VN}
		\frac{d\rho(t)}{dt}=-i\,\mathcal{L}\,\rho(t),  
	\end{equation}
	where $\mathcal{L}$ is the Liouville superoperator, defined as
	\begin{equation}
		\mathcal{L}X=
		\begin{cases}
			i\{H,X\}, & \text{(classical)},\\[8pt]
			\displaystyle\frac{1}{\hbar}[H,X], & \text{(quantum)}.
		\end{cases}    
	\end{equation}
	Here, $\{\,\cdot\,,\,\cdot\,\}$ and $[\,\cdot\,,\,\cdot\,]$ denote the Poisson bracket and the commutator, respectively~\cite{nitzan2024chemical}. This formulation allows us to treat classical and quantum dynamics in a unified manner. With this choice one recovers the standard forms:
	\begin{equation}
		\frac{d\rho}{dt}=\{H,\rho\}\quad\text{(classical)},\quad
		\frac{d\rho}{dt}=-\frac{i}{\hbar}[H,\rho]\quad\text{(quantum)}.   
	\end{equation}
	For any dynamical observable $\mathcal{O}$ (no explicit time dependence),
	\begin{equation}
		\frac{d\mathcal{O}}{dt}=i\,\mathcal{L}\,\mathcal{O},   
	\end{equation}
	so the formal time evolution is
	\begin{equation}
		\mathcal{O}(t)=e^{\,i\mathcal{L}(t-t_0)}\mathcal{O}(t_0);\; \qquad
		\rho(t)=e^{-\,i\mathcal{L}(t-t_0)}\rho(t_0).    
	\end{equation}
	Since the Liouvillian is Hermitian (one can show $\mathcal{L}=\mathcal{L}^\dagger$), the operator $e^{\pm i\mathcal{L}t}$ serves as the time evolution operator~\cite{kubo1983book}. We now investigate the evolution of the density operator under an external perturbation in the linear regime. The Liouvillian can be split as $\mathcal{L}=\mathcal{L}_0+\mathcal{L}_{\text{pert}}$, associated with $H_0$ and $H'$ respectively. The perturbed density operator is written as $\rho(t)=\rho_{\text{eq}}+\delta\rho(t)$. Inserting these into Eq.~\eqref{VN}, we obtain~\cite{balakrishnan2020elements, zwanzig2001nonequilibrium}
	\begin{equation}
		\frac{d\delta\rho(t)}{dt} = -i\mathcal{L}_{\text{pert}}\rho_{\text{eq}} - i\mathcal{L}_0\delta\rho(t),
	\end{equation}
	where we neglect the higher-order term $i\mathcal{L}_{\text{pert}}\delta\rho(t)$. Also, we set $\frac{d\rho_{\text{eq}}}{dt}=0$ and $i\mathcal{L}_0\rho_{\text{eq}}=0$, since $\rho_{\text{eq}}$ is stationary. Solving this equation, we obtain
	\begin{equation}\label{del_rho_unified}
		\delta \rho(t) = \int_{-\infty}^t \lambda(t^\prime)\,\mathcal{C}[A(t^\prime - t),\rho_{eq}]\,dt^\prime,
	\end{equation}
	where the operation $\mathcal{C}[\cdot,\cdot]$ is defined as
	\begin{equation}
		\mathcal{C}[X,Y] =
		\begin{cases}
			\dfrac{i}{\hbar}[X,Y], & \text{(quantum)}, \\[1em]
			\{X,Y\}, & \text{(classical)}.
		\end{cases}
	\end{equation}
	
	%\begin{equation}
	%\frac{d(\rho_{eq}+\delta\rho(t))}{dt}=-i\mathcal{L}_0\rho_{eq}-i\mathcal{L}_{pert}\rho_{eq}-i\mathcal{L}_0\delta\rho(t)-i\mathcal{L}_{pert}\delta\rho(t)
	%\end{equation}
	
	%We can solve the above differential equation using an integration factor $e^{i\mathcal{L}_0 t}$ and set $\delta \rho(t)\rightarrow 0$ at the initial condition ($t\rightarrow-\infty$). So, the following steps are as follows. 
	%\begin{equation}\label{del rho quantum}
	%\begin{split}
	%\int_{-\infty}^t d(e^{i\mathcal{L}_0t}\delta \rho(t))=& \int_{-\infty}^t-ie^{i\mathcal{L}_0 t^\prime }\mathcal{L}_{pert}(t^\prime)\rho_{eq}d t^\prime                             \\
	%\text{or,}\quad      \delta \rho(t)=& \int_{-\infty}^t -ie^{i\mathcal{L}_0(t^\prime-t)} \mathcal{L}_{pert}(t^\prime)\rho_{eq}dt^\prime\\
	%\text{or,}\quad \delta \rho(t)=& \int_{-\infty}^t -\frac{i}{\hbar}e^{i\mathcal{L}_0(t^\prime-t)} [H^\prime(t^\prime),\rho_{eq}]dt^\prime\\
	%=&  \frac{i}{\hbar}\int_{-\infty}^t F(t^\prime)e^{i\mathcal{L}_0(t^\prime-t)} [A(0),\rho_{eq}]dt^\prime\\
	%\text{or,}\quad \delta \rho(t) =&  \frac{i}{\hbar}\int_{-\infty}^t F(t^\prime) [A(t^\prime-t),\rho_{eq}]dt^\prime\\
	%\end{split}
	%\end{equation}
	
	%The time evolution here is only due to the unperturbed Liouvillean ($\mathcal{L}_0$). 

	From now on, we will focus on the quantum case, since the classical results of LRT can be directly obtained by replacing commutators with Poisson brackets, i.e.,
	$\frac{i}{\hbar}[\,\cdot,\cdot\,] \;\longleftrightarrow\; \{\cdot,\cdot\}$.
	For an arbitrary observable $B$, its expectation value in the perturbed state is $
	\langle B\rangle_{\rho(t)}=\langle B\rangle_{\rho_{eq}}+\langle B\rangle_{\delta\rho(t)}$.
	The equilibrium part $\langle B\rangle_{\rho_{eq}}$ is given in Eq.~\eqref{eqm_B}. Using Eq.~\eqref{del_rho_unified} and the definition of the average, the perturbative correction is calculated as
	\begin{equation}\label{change in B}
		\langle B\rangle_{\delta\rho(t)} = \operatorname{Tr}\{\delta\rho(t)\,B(0)\} = \int_{-\infty}^t dt^\prime \lambda(t^\prime) \phi_{AB}(t-t^\prime)
	\end{equation}
	where, the \textit{response function} is defined as~\cite{kubo1957statistical,kubo1983book}  \begin{equation}\label{phi-AB}
		\phi_{AB}(t-t^{\prime}) := \frac{1}{i\hbar}\operatorname{Tr}\!\left\{\rho_{\rm eq}[A(t^{\prime}), B(t)]\right\},
	\end{equation}  
	and depends only on the equilibrium state. This response of the observable $B$ due to external drive $\lambda(t)$ through the observable $A$, is \textit{linear, causal}, and \textit{retarded} \\ ~\cite{balakrishnan2020mathematical}.

	For a relaxation measurement, the system is first driven by an external force from $t=-\infty$ to $t=0$. When the force is suddenly switched off, any perturbed observable $B$ relaxes back to equilibrium as  
	\begin{equation}
		\begin{split}
			\langle B\rangle_{\delta\rho(t)} &= \int_{-\infty}^0 dt^\prime\,\lambda(t^\prime)\,\phi_{AB}(t-t^\prime) \\
			&= \lambda \int_{-\infty}^t ds\,\phi_{AB}(s), \quad t>0.
		\end{split}     
	\end{equation}  
	The \textit{relaxation function} is then defined as~\cite{kubo1957statistical,kubo1983book}
	\[
	\Phi_{AB}(t) = \lim_{\epsilon\to 0}\int_{-\infty}^t \phi_{AB}(s)e^{-\epsilon s}\,ds,
	\]  
	which characterizes how $B$ relaxes after the perturbation is removed. Using Kubo’s identity for the canonical ensemble~\cite{nitzan2024chemical,bellac2004equilibrium},  
	\[
	[A,\rho_{\rm eq}] = e^{-\beta H_0}\int_0^\beta e^{-\alpha H_0}[H_0,A]e^{-\alpha H_0}\,d\alpha,
	\]  
	(where $\alpha$ is a dummy variable), the relaxation function can be written as  
\begin{equation}
\begin{split}
\Phi_{AB}(t-t^\prime) &= 
\int_0^\beta 
\operatorname{Tr}\!\big(\rho_{\rm eq} A(-i\hbar\alpha) B(t-t^\prime)\big) \, d\alpha \\
&\quad - \int_0^\beta 
\operatorname{Tr}(\rho_{\rm eq} A)\operatorname{Tr}(\rho_{\rm eq} B) \, d\alpha
\end{split}
\end{equation}

	This function is also referred to as the \textit{canonical correlation function}. However, in quantum mechanics, it is often more useful to express the \textit{relaxation function} in a form that accounts for the non-commutativity of operators, leading to the \textit{symmetrised correlation function}, denoted as $\Psi_{AB}(t-t^\prime)$~\cite{kubo1983book}. Since it accurately captures the response (i.e., the fluctuation from equilibrium) caused by an external disturbance through time correlations between two dynamical observables, LRT has become increasingly important in \textit{quantum thermodynamics}, where the control~\cite{gelbwaser2015thermodynamics} and optimisation of quantum devices governed by fluctuations~\cite{das2024fluctuations}, particularly quantum fluctuations, are of growing interest. In our context, the \textit{autocorrelation function} ($A = B$), which measures how an observable correlates with itself at different times, is of central importance and will be discussed in the next section.

	%These functions are derived here has huge implications in the field of fluid dynamics, transport phenomena (classical and quantum both), condensed matter physics, chemical kinetics and many more.

	\subsubsection{Quanum Work Fluctuations and LRT:}

	We live in an era where we aim to reach, utilise, and design systems on scales that have never been achieved before~\cite{rossnagel2016single,myers2022quantum,gupt2022PRE,sur2023quantum}. To fully understand and optimise machines at the quantum level, where thermal and quantum fluctuations play equally important roles, we need to deepen our theoretical understanding of FTs, a.k.a.,  \textit{quantum fluctuation theorems} (QFTs)~\cite{hanggitalkner2015theotherqft,funo2018quantumft,strasberg2022quantum,potts2024quantumthermodynamics}. While QFTs can be formulated for various thermodynamic quantities, in this article we focus exclusively on work. Within the LRT framework, formulating QFTs for dissipative work has yielded new insights~\cite{andrieux2008quantum}. More recently, Guarnieri \textit{et al.}~\cite{guarnieri2024generalized} demonstrated that even near equilibrium, the work distribution is strongly influenced by quantum fluctuations, which are captured through the relaxation functions.

	%We are in an era where we want to reach, utilise, and architect in such scales that humankind has never reached before. To fully understand and optimise machines at the quantum level, where thermal fluctuations play a crucial role and an additional source of quantum fluctuation is also there, we have to increase our theoretical understanding of the FTs there, also known as \textit{quantum fluctuation theorems}(QFTs). This \textit{other QFT} can be established and investigated for many thermodynamic parameters like entropy, heat, but in this article, we will restrict ourselves to work. Developing QFTs for work in the framework of LRT gave some new, insightful results. Guarnieri et al. showed in a recent paper that even in the near-equilibrium regime, the work distribution is profoundly affected by the quantum fluctuations with the help of \textit{ relaxation functions}, which we introduced in the earlier section. 

	For example, consider a quantum system with Hamiltonian $H_0$, initially thermalised in the normalised Gibbs state $\rho_{\text{eq}} = e^{-\beta H_0}$. The system is then perturbed unitarily as $H_t = H_0 - \lambda(t)A$ over $t \in [0,\tau]$, where $A$ is an observable and $\lambda(t)$ is a weak dimensionless driving protocol ($\lambda(t)\ll1$, $\|A\|=1$). At $t=\tau$, when the drive is switched off, the relaxation of $A(\tau)$ is captured by the \textit{symmetrized autocorrelation function} $\Psi_0(=\Psi_{AA})(t-t')$, which encodes correlations of $A$ at different times. Since changes in $A$ also modify the Hamiltonian, $\Psi_0$ links correlations between $H_\tau$ and $H_0$, providing a natural connection to the very definition of \textit{quantum work}~\cite{guarnieri2024generalized,talkner2007fluctuation}.

	The most accomplished and widely used definition of work in QTD is the \textit{two-time projective measurement}~\cite{strasberg2022quantum,deffner2019quantumthermo} (TPM) scheme, where the system’s energy is measured at the beginning and end of the protocol. At $t=0$, a projective energy measurement is performed, then the system evolves under the time-dependent Hamiltonian, and at $t=\tau$ a second measurement is made. In this framework, work $W$ is a stochastic variable. For a single realization, it is given by $
	W[|m\rangle;|n\rangle] = E_n(\lambda(\tau)) - E_m(\lambda(0))$, 
	where $|m\rangle$ and $|n\rangle$ are the initial and final eigenstates with energies $E_m(\lambda(0))$ and $E_n(\lambda(\tau))$, respectively. The probability distribution of work is obtained by averaging over an ensemble of realizations~\cite{esposito2009nonequilibrium} 
	\begin{equation}\label{TPM1}
		\mathcal{P}(W) = \sum_{W=E_n-E_m} \delta\left(W - W[|m\rangle;|n\rangle]\right)\, 
		p(|m\rangle \rightarrow |n\rangle),
	\end{equation}  
	where $p(|m\rangle \rightarrow |n\rangle)$ denotes the probability of observing the transition $|m\rangle \rightarrow |n\rangle$. If the system is initially in a thermal Gibbs state $\rho_{eq}$, Eq.~\eqref{TPM1} reduces, as shown by Scandi~\textit{et al.}~\cite{scandi2020quantum}, to the standard TPM definition of work:  
	\begin{equation}\label{TPM2standard}
    \begin{split}
        &\mathcal{P}(W) \\ =& \sum_{W=E_n-E_m} 
		\langle m(\lambda(0))|\rho_{eq}|m(\lambda(0))\rangle 
		\, \big|\langle m(\lambda(0))|n(\lambda(\tau))\rangle\big|^2.
        \end{split}
	\end{equation}  
	This expression represents the product of the probability of the system initially being in state $|m\rangle$ with the transition probability $|m\rangle \rightarrow |n\rangle$. However, since each projective energy measurement perturbs the system --- and may even destroy its state --- the full $\mathcal{P}(W)$ cannot be directly accessed in practice through this scheme~\cite{scandi2020quantum,deffner2019quantumthermo}.

	%We can rewrite Eq.~\eqref{TPM1} when the system initially is in thermal Gibbs state ($\rho_{eq}$), as Scandi et al. have shown, as a standard definition of TPM work as
	%\begin{equation}\label{TPM 2 standard}
	%p(W)=\sum_{W=E_m-E_n} \langle m(\lambda(0))|\rho_{eq}|m(\lambda(0))\rangle|\langle m(\lambda(0))|n(\lambda(\tau))\rangle|^2
	%\end{equation}
	%So, this basically represents the product of the probability of the state initially in the $|m\rangle$ state with the transition probability of $|m\rangle\rightarrow|n\rangle$. As every time we measure the energy of the system, the system gets perturbed, can even be destroyed, we can not get the full distribution $W$ in this way.

	\indent However, from the statistical theory, we know that any probability distribution can be reconstructed from its \textit{cumulant generating function} (CGF)~\cite{strasberg2022quantum}, defined as
	\begin{equation}\label{CGF}
		K(\lambda)=\ln{\langle e^{-\lambda\beta W }\rangle}
		:=\ln{\int_{-\infty}^\infty}d W \mathcal{P}(W)e^{-\beta\lambda W},
	\end{equation}
	where $\lambda=\lambda(\tau)$, denotes the final value attained by the protocol. The CGF is additive under independent random measurements, and allows one to recover the probability distribution $\mathcal{P}(W)$ through an appropriate inverse Fourier transform. Moreover, all the statistical information of $\mathcal{P}(W)$, such as cumulants, can be computed by differentiating the CGF~\cite{kardar2007book}: 
	\begin{equation}\label{cumulants}
		\kappa_W^{(n)}=(-\beta)^{-n}\frac{d^n}{d\lambda^n}K(\lambda)\bigg{|}_{\lambda=0}. 
	\end{equation}

	Since $W = \Delta F + W_{\text{diss}}$, the CGF naturally separates into two parts: a deterministic term depending only on the free-energy difference, and dissipative contribution due to the irreversible process, as shown by Scandi \textit{et al.}~\cite{scandi2020quantum}
	\begin{equation}\label{diss cgf}
		K(\lambda)=-\beta \lambda\Delta F+K^{\text{diss}}(\lambda),
	\end{equation}
	where $K^{\text{diss}}(\lambda)=\ln{\int_{-\infty}^\infty}d W_{\text{diss}} \mathcal{P}(W_{\text{diss}})e^{-\beta\lambda W_{\text{diss}}}$. So, the most interesting quantity in QTD is the \textit{dissipated work}, $W_{\text{diss}}$, which quantifies the irreversible cost of driving the system out of equilibrium. Its cumulants encode the full statistical information relevant to QFTs.

	\indent Guarnieri \textit{et. al.}~\cite{guarnieri2024generalized}, building on the elegant method of Shitara and Ueda~\cite{shitara2016determining}, established a landmark bridge between LRT and QFTs as
	\begin{equation}\label{LRT-CGF}
		K^{\text{diss}}(\lambda)=-\int_0^\tau dt \int_0^\tau dt^\prime \dot{\lambda}(t)\dot{\lambda}(t^\prime)[g_\lambda*\Psi_0](t-t^\prime),
	\end{equation}
	where a function of $\lambda$, $g(\lambda)$, has to be convoluted with our old friend $\Psi_0$ to get the CGF of the dissipated work. The symmetry of $g(\lambda)=g(1-\lambda)$ ensures $K^{\text{diss}}(\lambda)=K^{\text{diss}}(1-\lambda)$~\cite{guarnieri2024generalized}. Operating an inverse Laplace transform on it, gives our pre-acknowledged companion
	\begin{equation}\label{EW of diss W}
		\frac{\mathcal{P}(W_{\text{diss}})}{\mathcal{P}(-W_{\text{diss}})}=e^{\beta W_{\text{diss}}},
	\end{equation}
	a special form of ESFT~\cite{guarnieri2024generalized}, now established in the quantum regime. The essential finding of Eq.~\eqref{LRT-CGF} is that any phenomenological relaxation function can fully determine quantum work distribution in the linear regime. Moreover, this framework predicts all higher-order cumulants are positive, i.e., $\kappa_W^{(n)}\ge0$~\cite{guarnieri2024generalized}, leading to non-Gaussian tail (such as skewness and excess kurtosis), driven by quantum fluctuations~\cite{scandi2020quantum}. To quantify these non-classical effects, Guarnieri \textit{et al.}~\cite{guarnieri2024generalized} derived a bound stronger than standard \textit{thermodynamic uncertainty relation} (TUR)~\cite{agarwalla2018assessing,seifert2025stochastic}. In LRT, the the conventional TUR reads (in terms of the Fano factor)~\cite{guarnieri2024generalized}:
	\begin{equation}\label{fano}
		F_W=\frac{\sigma^2_{W_{diss}}}{\langle W_{diss}\rangle}\ge 2 k_B T.
	\end{equation}

	While, Guarnieri \textit{et al.}~\cite{guarnieri2024generalized} have rigorously proved that by defining a normalised probability distribution function $\widetilde{P}(\omega)$, associated with the system relaxation function $\widetilde{\Psi}_{0}(\omega)$, over the \textit{pseudomodes} frequencies $\omega\in [0,\infty)$, the Fano factor can be expressed as: \begin{equation}\label{fano quantum}
		\begin{split}
			F_W
			&=~\langle \hbar\omega \coth(\beta\hbar\omega/2) \rangle_{\widetilde{P}}
			= 2\langle \mathcal{E}_\omega \rangle_{\widetilde{P}} \\
			&\ge~\hbar \langle \omega \rangle_{\widetilde{P}}
			\coth\!\left(\beta \hbar \langle \omega \rangle_{\widetilde{P}}/2\right)
		\end{split}
	\end{equation}
	
 Equation~\eqref{fano quantum} provides a stronger bound than Eq.~\eqref{fano}. Although one expects $F_W$ to saturate to classical behaviour in the long-time limit, $\mathcal{E}_\omega$ --- the average energy of a \textit{quantum harmonic oscillator} --- includes the contribution of its average \textit{zero-point energy}, which persists even in the vacuum state and prevents the system from becoming fully classical. At short times, however, \textit{vacuum fluctuations} give rise to pronounced quantum signatures in the work distribution for various phenomenological Brownian-motion models.
	
\section{The way from here}

\begin{framed}
\noindent
``The first quantum revolution gave us new rules that govern physical
reality. The second quantum revolution will take these rules and use
them to develop new technologies.''

\vspace{2pt}
\hfill --- Jonathan P. Dowling~\cite{dowling2003quantum}
\end{framed}

%\begin{center}
%\begin{minipage}{\textwidth}
%\begin{tcolorbox}[colback=gray!10, colframe=black!50, width=0.48\textwidth, boxrule=0.5pt, arc=3mm]
%\itshape
%“The first quantum revolution gave us new rules that govern physical reality. The second quantum revolution will take these rules and use them to develop new technologies and offer the next level of opportunities.”

%\hfill --- Ian R. McAndrew
%\end{tcolorbox}
%\end{minipage}
%\end{center}

Throughout this article, we have outlined a trajectory from 19th-century thermodynamics to modern formalisms, passing through the emergence of stochastic thermodynamics and culminating in the \textit{rise of quantum thermodynamics}~\cite{halpern2022quantum}, formulated here within the language of LRT near equilibrium. In this framework, the QFT for TPM-based dissipated work reaches a height without relying on slow-driving or weak coupling aproximations~\cite{guarnieri2024generalized}. This perspective may shed light on non-Markovian effects and work fluctuations across phase transitions~\cite{campbell2025roadmap}, while remaining experimentally relevant since the Fano factor is directly measurable~\cite{guarnieri2024generalized}.

\indent While the TPM definition is useful, \textit{quantum work} is not an observable~\cite{talkner2007fluctuation}, and the approach suffers from serious drawbacks~\cite{deffner2019quantumthermo}. In particular, it neglects measurement back-action, and projective energy measurements are neither feasible nor practical~\cite{deffner2019quantumthermo}. These limitations motivate describing \textit{work fluctuations} in terms of variations of physical quantities (VPQs)~\cite{silva2025fundamental}. A promising alternative is the \textit{two-time observable protocol}, which is experimentally accessible in platforms such as trapped ions, plasma drops, superconducting qubits, and NMR setups~\cite{silva2025fundamental}. In this context, linear response theory may offer a natural framework for modeling these systems.

%\indent We can see \textit{quantum work} is not an observable, but the TPM definition, although useful but has some serious drawbacks. One issue is that there is negligence in considering back-action on the system due to measurements, and another is that the experimentally set-up for the projective measurements of the energy is neither feasible nor practical. These limitations in TPM add fuel to the motivation to express and analyse the \textit{work fluctuations} more precisely in the \textit{variation} of physical quantities (VPQs). There is a growing interest in \textit{two times observable protocol} which can replace other definitions in the near future and can be more experimentally matched up for various systems like trapped ion, plasma drop, superconducting qubits, and NMR systems. If we can express and model these setups, LRT can be the key tool for developing the theoretical background. 

\indent As our knowledge advances, we are steadily moving toward the ability to design highly efficient, optimised devices---such as quantum computers, sensors, and metrological tools --- that harness quantum phenomena as fundamental resources. This heralds the onset of the \textit{second quantum revolution}~\cite{dowling2003quantum}, in which quantum effects extend beyond atoms and molecules to profoundly shape technology, society, and civilisation.

\section{Acknowledgements}
We acknowledge IIT Kanpur for providing partial financial support.

\end{document}